\documentclass[preprint,12pt]{elsarticle}
\journal{Nuclear Instruments and Methods A}

\usepackage{hyperref}
\usepackage{graphicx}
\usepackage{xcolor}
\usepackage{amssymb,amsmath,bm}
\usepackage{subfig}
\usepackage{booktabs}
\usepackage{xspace}
\usepackage[version=4]{mhchem} %Typeset chemical formulas
\usepackage[siunitx]{circuitikz}
\usetikzlibrary{shapes.geometric, arrows,calc,positioning}
\usepackage{textcomp}
\usepackage{gensymb}
\usepackage[T1]{fontenc}

\begin{document}
\begin{frontmatter}

\title{NA62 Liquid Krypton Purity Monitor}
\author[1] {J. Bremer}
\affiliation[1]{ CERN,European Organization for Nuclear Research, Espl. des Particules 1, 1211 Meyrin  Switzerland}
\author[2,3]{ D. Bryman}
\affiliation[2]{Department of Physics and Astronomy, University of British Columbia 6224 Agricultural Road, Vancouver V6T 1Z1 Canada}
\affiliation[3]{TRIUMF, 4004 Wesbrook Mall, Vancouver V6T 2A3 Canada}
\author[1]{H. Danielsson}
\author[4,5]{V. Falaleev}
\affiliation [4] { Joint Institute for Nuclear Research, 141980 Dubna (MO), Russia}
\affiliation[5]{INFN, Sezione di Perugia, I-06100 Perugia, Italy}
\author[1]{T. Koettig}
\author[3]{L. Kurchaninov}
\author[1]{J. Liberadzka-Porret}
\author [1] {A. Onufrena}
\author[3]{B. Velghe}

\date{October 10, 2022}
\newpage
\begin{abstract}
    A system for determining  the purity of liquid krypton employed in the NA62 rare kaon decay experiment at CERN  was developed based on the use of a time projection chamber. The attenuation of drifting ionization electrons from  absorption of 511 keV gamma rays in liquid krypton was measured to estimate the purity. The setup was tested with krypton purified from commercial sources. 
\end{abstract}
\end{frontmatter}

\newpage
\section{Introduction}
The liquid krypton (LKr) ionization calorimeter (LKRC) plays a key role in the photon veto and particle identification systems for rare kaon decay measurements done at the CERN experiment NA62~\cite{NA62:2021zjw}.
Detailed descriptions of the LKRC can be found in \cite{NA62:2017rwk, NA48:2007mvn}.
The LKRC active volume is longitudinally segmented in 127~cm-long beryllium-copper ribbons.
These ribbons form an octagonal grid of 13,248~$1~\times~2$~cm\textsuperscript{2} individual towers facing the beam direction. 
The cylindrical cryo-vessel contains about 9~m\textsuperscript{3} of LKr. 
Boil-off gas passes through filters before being re-condensed by an argon cooler, ensuring continuous recirculation. 

Due to small and transitory leaks in the LKRC system during the period 2016--17, it became necessary to replenish the krypton supply using commercially obtained gas which is typically available with purity at the ppm level for several contaminants including \ce{O2} and \ce{H2O}.
However, for operation in the LKRC, purification of the krypton gas at the (\ce{O2} equivalent) ppb level is required to restrict charge losses to a negligible level.
Measurement of the attenuation (or effective lifetime) of drifting ionization\cite{Manenti:2020gzi} is desirable before it is transferred to the LKRC.

The principal electro-negative impurities in noble liquid drift chamber systems are usually \ce{O2} and \ce{H2O}.
During the development and commissioning phase of the LKRC, the maximum electron drift time was measured under normal operating conditions (with 3~kV applied to the anodes) to be 3~{\textmu}s~\cite{NA48:1995kde}.
The ionization electron lifetime was found be $>300$~{\textmu}s~\cite{NA48:2007mvn} which is sufficient for NA62 operations. 

To meet NA62 LKRC requirements, a purity monitoring system was developed with the goal of confirming electron lifetimes $\geq 300$~{\textmu}s.
The system is based on the use of a small (32~cm\textsuperscript{3}) time projection chamber (TPC) installed in a cryogenic vessel containing about 2~L of LKr.
The device allows measurement of the attenuation of ionization charge as a function of the drift time and applied electric field.

This report describes the cryogenic and purity measurement systems, and the purity measurements made with newly purified krypton. The following sections cover the cryogenic system (Section~\ref{cryo}), the detection apparatus (Section~\ref{sec:cell}), and the measurements performed (Section~\ref{Measurements}).

\section{Purification and Cryogenic Systems}
\label{cryo}
\begin{figure}[ht]
\centering
\includegraphics[width=1.\textwidth]{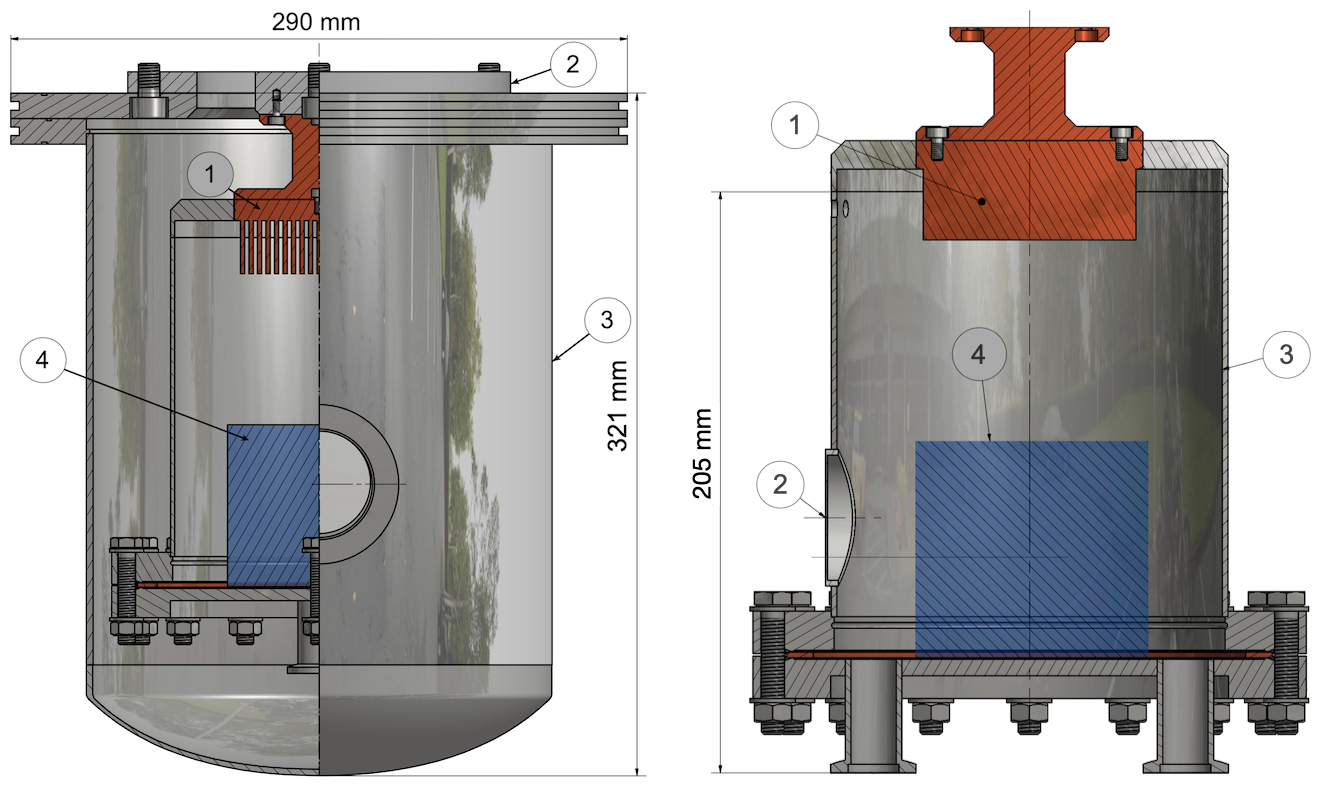}
\caption{Drawing of the vacuum vessel (cutaway) and LKr vessel housing the measurement chamber. Left: 1)~Heat exchanger (HEX) attached to the cryocooler cold head (not shown); 2)~Vacuum flange interface to the cryocooler; 3)~Outer vacuum vessel; and 4)~TPC. Right: 1)~HEX; 2)~radiation window (1~mm thick stainless steel); 3)~LKr vessel; and 4)~TPC.}
\label{vessel}
\end{figure}
The cryogenic setup and the procedures used to purify and condense LKr are described in~\cite{Liberadzka-Porret:2022nje}.
Briefly, filtered krypton gas was condensed inside the LKr vessel with a heat exchanger (HEX) linked to a cryocooler  and maintained at a temperature of 119.8~K. 
The filters used to purify commercially supplied krypton gas were the same as in the LKRC recirculation system: two Sertronic type N gas purifiers\footnote{http://www.sertronic.com/gas-purifiers/sertronic-purifiers/neutral-gas-purifier/} in series.
In the filters, the feed gas passes through a special catalyst which traps oxygen, while water and carbon dioxide are adsorbed on a molecular sieve bed.
The principal impurities removed are \ce{O2}, \ce{CO}, \ce{CO2}, and \ce{H2O}. 
A single stage coaxial pulse tube cryocooler was used as a cooling source to liquefy the krypton.
The cooling power of the cryocooler (with a 3~kW compressor) at 120~K is approximately 55~W.
The temperature of the cold head was controlled at 119~K during condensation and then at 119.8~K during stable operation.
Figure~\ref{vessel} shows an assembly drawing of the vacuum insulated LKr vessel with the TPC, HEX, and the radiation entrance window.
The inner LKr vessel cryostat is covered by 20 layers of multi-layer insulation.
Prior to the measurements described in Section~\ref{Measurements}, the LKr vessel was baked at 60~{\textdegree}C for 3~weeks by running a current through a temporarily applied heating tape\footnote{The HEX cold head of the cryocooler cannot be heated above 60~{\textdegree}C and the MPPCs described in section \ref{sec:cell} cannot be heated to more than 80~{\textdegree}C.} and pumping vacuum to a level below $2\times 10^{-7}$~mbar.
Since residual contamination due to the cryostat, TPC, and gas handling apparatus could not be determined, the attenuation measurements discussed below can only be considered as upper limits.
It is also assumed that the results are unaffected by evaporation and condensing cycles.

\section{Detection Apparatus}
\label{sec:cell}
The TPC detector system, previously used in the measurements with liquid xenon~\cite{Amaudruz:2009vs}, was modified for measurements using LKr.
As discussed below, avalanche photodiodes were replaced by Multi-Pixel Photon Counters (MPPC) to provide triggering for the ionization drift time measurement and preamplifiers were installed on the anodes. 

\begin{figure}[!ht]
\includegraphics[width=\textwidth]{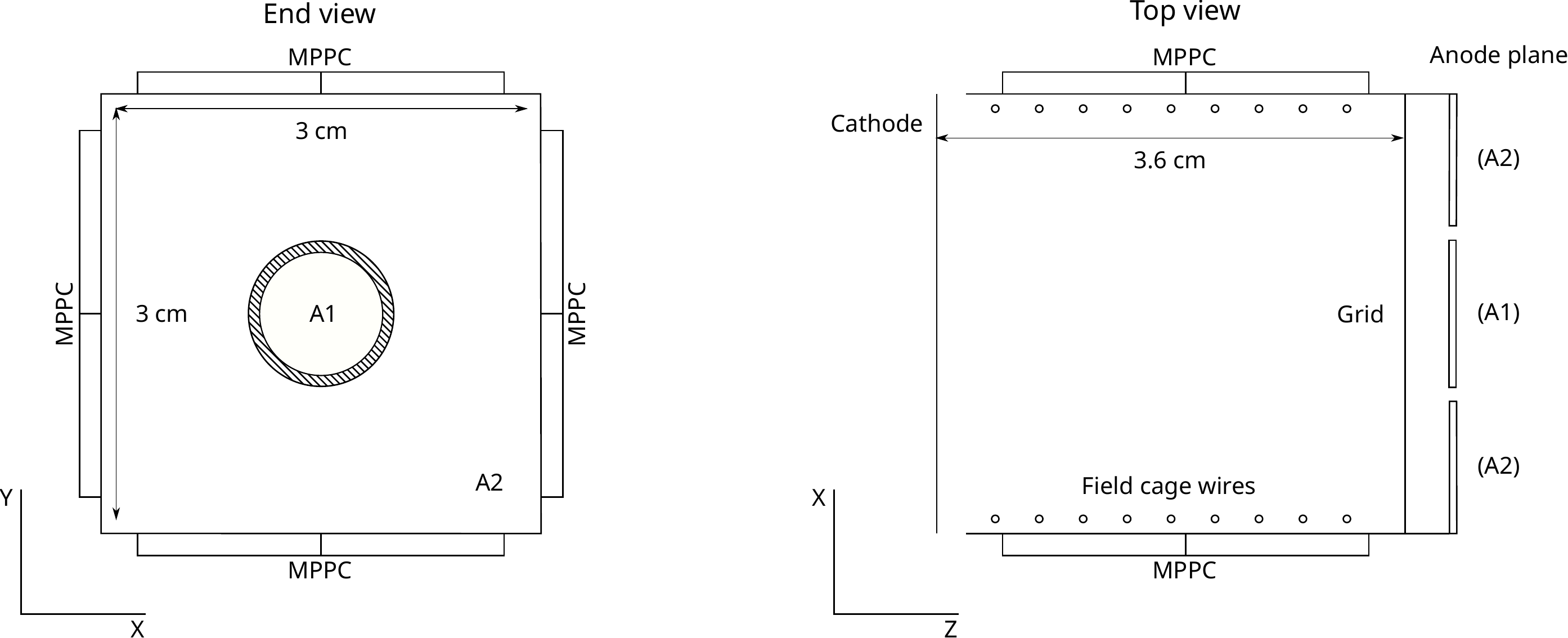}
\caption{Schematic of the TPC. Left: End view showing the central (A1) and outer (A2) anodes and locations of the MPPC light sensors. Right: Side view showing the drift region with anodes, grid, cathode, and field cage wires; 511 keV gamma rays from annihilation of positrons emitted by a \textsuperscript{22}Na source enter through the cathode from the left.}
\label{TPC0}
\end{figure}

The cryostat was filled with LKr immersing the TPC which was located approximately 1~cm from the cryostat entrance window (see Figure~\ref{vessel}).
As shown in Figure~\ref{TPC0}, the TPC features ground potential anodes, a shielding grid (25~{\textmu}m dia. wires, spaced 3~mm apart) separated from the anodes by 4.8~mm, a field cage, and a negatively biased cathode plane with a $3 \times 3 \times 3.6$~cm\textsuperscript{3} drift volume; the grid shielded the anode from induced charge during the ionization drift time.
Charge was collected on a central 1~cm dia. electrode (A1) used to select centrally located events and on a $3~\times~3$~cm\textsuperscript{2} outer electrode (A2).
An electric drift field up to 1.5~kV/cm could be applied between the cathode and the shielding grid which were separated by 3.6~cm.
The electric field between the anodes and grid was maintained at twice the drift field value to assure maximum transmission~\cite{Bunemann}.
A field cage consisted of nine wires strung around four ceramic pillars in the corners of the chamber; the wires were spaced by 3~mm with the first of these 3~mm from the shielding grid.
The voltage was distributed by 100~M$\Omega$ resistors located outside the cryostat.
The cathode was separated by 9.8~mm from the last field cage wire; it consisted of a 200~{\textmu}m thick stainless steel plate with a pattern of holes to facilitate the flow of LKr into the TPC.

\begin{figure}
\includegraphics[width=\textwidth]{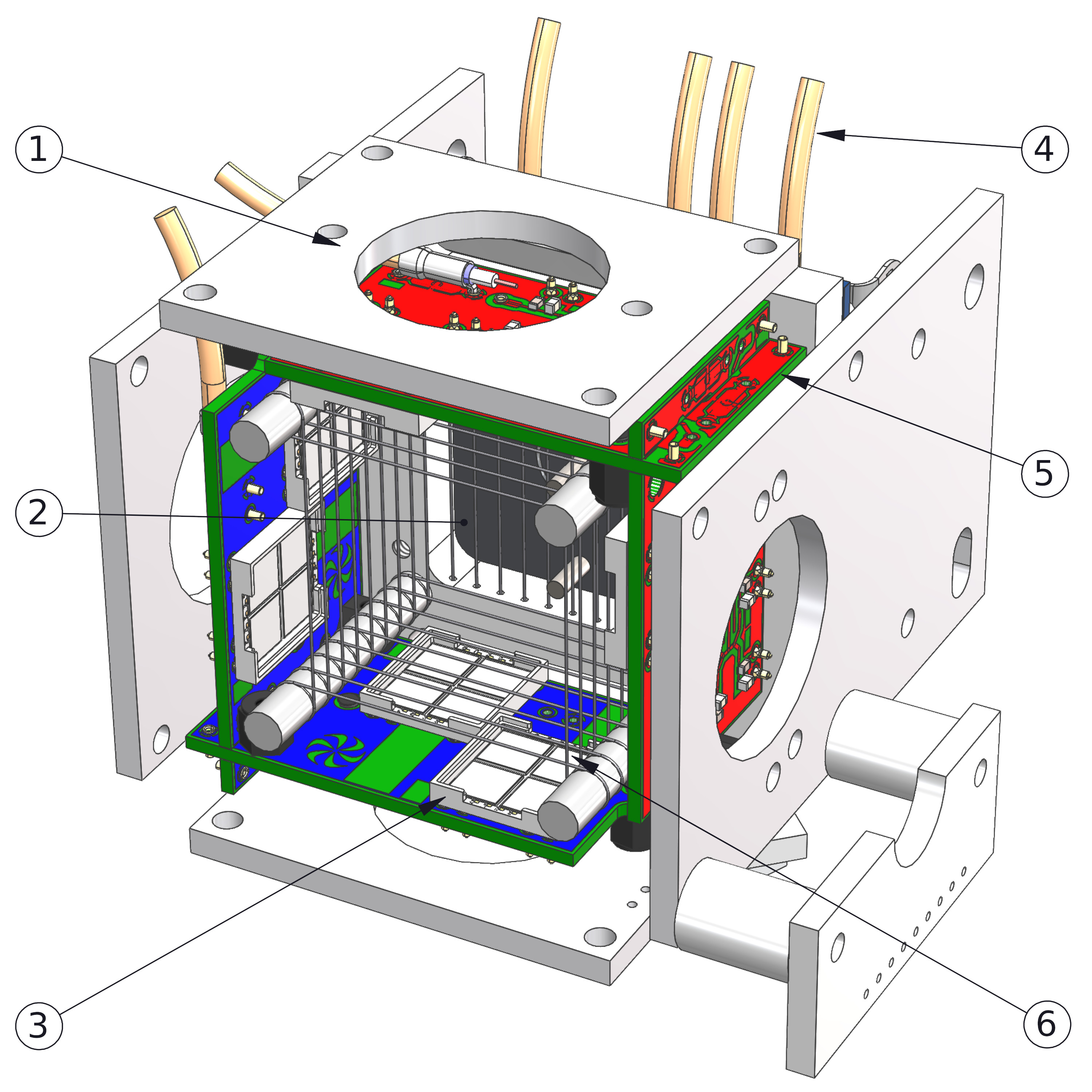
}
\caption{3D model of the TPC with the cathode plane removed for clarity. Components shown are 1) Macor{\textregistered} structural panels, 2) anodes and grid, 3) MPPCs, 4) coaxial signal and power cables, 5) support PCB for the MPPCs, and 6) field cage wires.}
\label{TPC1}
\end{figure}

As indicated in Figure~\ref{TPC1}, eight windowless Hamamatsu VUV4-MPPC\footnote{Hamamatsu model S13371-6050CN-02 VUV Multi-Pixel Photon Counter.} were assembled to view the prompt 150~nm scintillation light from LKr~\cite{Aprile:2008bga} through the field cage wires. The signals from two groups of four MPPCs were summed and sent to external room temperature post-amplifiers (MPPC1 and MPPC2).
Prior to the installation of the MPPC light sensors, the TPC was washed with isopropanol in an ultrasonic cleaner at a maximum temperature of 60~{\textdegree}C.
The completed detector system was then installed in the cryostat.

\begin{figure}
\includegraphics[width=\textwidth]{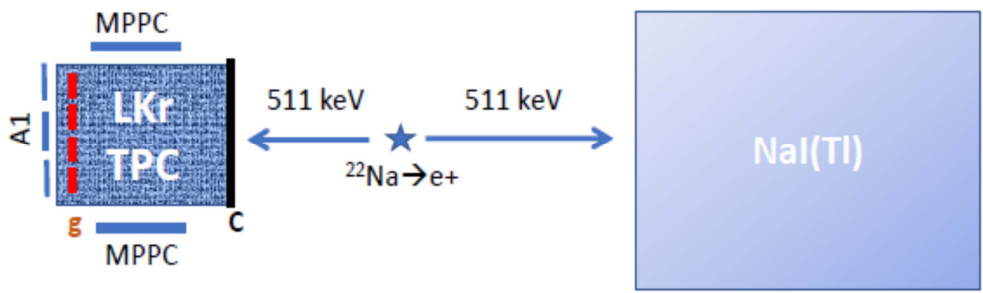}
\caption{Schematic diagram of the measurement system (not to scale). A \textsuperscript{22}Na positron annihilation source emits back-to-back 511~keV gamma rays detected by the TPC and a NaI(Tl) detector. The TPC central anode (A1), grid (g), cathode (C), and MPPCs are indicated.}
\label{layout}
\end{figure}

An external \textsuperscript{22}Na positron annihilation source (activity $2.5 \times 10^5$~Bq) was used to create energy deposits in the TPC for electron drift time measurements as indicated in Figure~\ref{layout}. 
511~keV photons produced from positron annihilation at the source were detected in the TPC in coincidence with an external NaI(Tl) crystal detector of dimensions 12.7~cm diameter and 12.7~cm in length.
The use of the coincidence with the NaI(Tl) detector significantly reduced the rate of random hits in the TPC due to the contamination of \textsuperscript{85}Kr as discussed in Section~\ref{Measurements} below.
A source collimator and the crystal were aligned in a brass support cylinder aimed at the central anode (A1) position previously determined relative to the cryostat entrance window shown in Figure~\ref{vessel}.
The coincidence between energy-selected 511 keV photoelectron events measured in the NaI(Tl) detector with light signals from the MPPCs provided the prompt trigger for the TPC ionization drift time measurements. 

Taking into account the energy needed to produce an electron--ion pair in LKr, $W = 20.5 \pm 1.5$~eV~\cite{Takahashi1974}, and assuming full charge collection efficiency and grid transparency at 0.77~kV/cm~\cite{Bunemann}, an A1 anode signal of about 2.5~fC was expected for a 511~keV photon undergoing absorption by photoelectron emission.
The anode signals were read out with adjacent N-Channel JFET (BF862) amplifiers situated in the LKr (see Figure~\ref{cold-preamp}) and connected by coaxial cables to room temperature post-amplifiers located outside the cryostat. 

\begin{figure}
\centering
\begin{circuitikz}
    \ctikzset{resistors/scale=0.8}
    \draw (0,0) node[ocirc,label=A1]{} -- ++(1,0) coordinate (n1) to [R=10M$\ohm$,*-] ++(0,-2) to ++(0,0) coordinate (gnd) node[tlground]{};
    \draw (n1) -- ++(0.5,0) node[njfet, anchor=G](J1){BF862};
    \draw (J1.D) -- ++(+2,0) node[ocirc,label=To feed-through]{};
    \draw (J1.S) -- (gnd -| J1.S) node[tlground]{};
\end{circuitikz}
\caption{Schematic of the cold preamps (see text).}
\label{cold-preamp}
\end{figure}
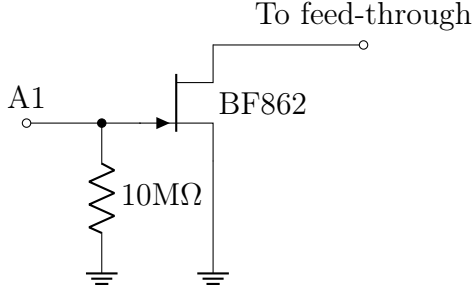

\begin{figure}
   \begin{tikzpicture}[scale=0.8, every node/.style={scale=0.7}]

   \tikzstyle{arrow} = [thick,->,>=stealth]
   \tikzstyle{thick_arrow} = [ultra thick,->,>=stealth]
   \tikzstyle{biarrow} = [thick,<->,>=stealth]
   \tikzstyle{box} = [rectangle, rounded corners, minimum width=3cm, minimum height=1.5cm,align=center, draw=black]
   \tikzstyle{amp_node_mask} = [rectangle, rounded corners, minimum width=3cm, minimum height=1.5cm,text centered]

\tikzset{
  pics/amp_node/.style args={#1}{
     code={
       \path (0,0) node[shape=rectangle,rounded corners,draw=black,minimum height=1.5cm,minimum width=3cm,text width=2.5cm,align=right] (sub_amp) {#1}
     ($(sub_amp.west)+(0.5,0)$) node[shape=isosceles triangle,draw=black,minimum height=1cm] (sub_amp) {};
     }
  }
}

   \path (0,0) node[box] (nai) {NaI(Tl) \& PMT}
        ++(0,-1) coordinate (anchor_top) {}
        ++(0,-1) pic{amp_node={MPPC1\\amplifier}} {} node[amp_node_mask] (mppc1_amp) {}
        ++(0,-1) coordinate (anchor) {}
        ++(0,-1) pic{amp_node={MPPC2\\amplifier}} {} node[amp_node_mask] (mppc2_amp) {}
        ++(0,-2) pic{amp_node={Anode post-amps}} {} node[amp_node_mask] (anode_amp) {};

  \path (anchor.center) -| ++(4.5,0) node[box] (trigger) {NIM trigger logic\\\& Fan-out}
  ++(4.5,0) node[box] (adc) {CAEN DT5725\\ADC}
  ++(4.5,0) node[box] (daq_pc) {DAQ PC};

    \draw [arrow] (nai.east) -- ($(nai.east)+(0.8,0)$) |- ($(trigger.west)+(0,0.4)$);
    \draw [arrow] (mppc1_amp.east) -- ($(mppc1_amp.east)+(0.4,0)$) |- ($(trigger.west)+(0,0.0)$);
    \draw [arrow] (mppc2_amp.east) -- ($(mppc2_amp.east)+(0.4,0)$) |- ($(trigger.west)+(0,-0.4)$);
    \draw [arrow] (anode_amp.east) -|  coordinate (C) (adc.south);

    \draw [arrow] let \p{1} = ($(anchor_top) - (trigger.north)$) in (trigger.north) -- ($(trigger.north)+(0,\y1)$) coordinate (A) -- ($(adc.north)+(0,\y1)$) coordinate (B)  -- (adc.north);

    \path (A) -- (B) node [midway, above] () {NaI, MPPC1, MPPC2 signals};
    \path (anode_amp.east) -- (C) node [midway, above] () {A1/A2 signals};
        
    \draw [biarrow] (trigger) -- (adc) node [midway,above] () {Trigger};
    \draw [arrow] (adc) -- (daq_pc) node [midway,above] () {Data};
    \path (trigger) -- (adc) node [midway,below] () {Control};
\end{tikzpicture}
\caption{Overview of the readout and data acquisition systems. The amplified MPPC and NaI(Tl) signals pass through NIM-based logic units that construct a trigger signal. Copies of these amplified signals are also sent to the ADC for digitization. The amplified anode signals (A1 and A2) connect directly to the ADC.}
\label{fig:readout_sys}
\end{figure}
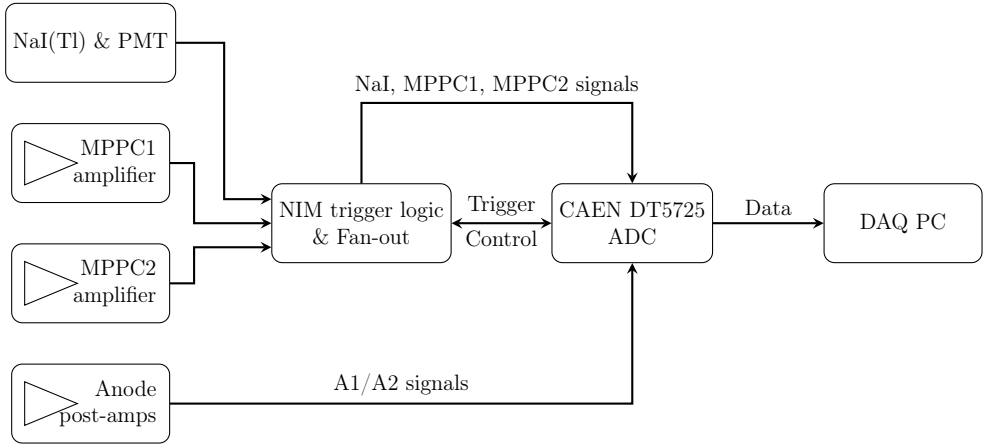
The readout system is outlined in Figure~\ref{fig:readout_sys}.
The signal from the NaI(Tl) detector was split to allow high and low level discriminators used to select the 511~keV photoelectron events.
The signals from the NaI(Tl) detector, MPPCs, and anodes were read out using a CAEN DT5725 14-bit 500~MS/s FADC. 
The FADC was connected through a USB port to a laptop computer running the MIDAS~\cite{MIDAS} data acquisition system. 

\section{Measurements}
\label{Measurements}
Initial measurements were performed at the CERN Cryolab and, after qualification, the apparatus was transported and reassembled adjacent to the LKRC in the NA62 experimental area.
Prior to measurements, the LKr vessel and connecting piping were baked-out as mentioned in Sec.~\ref{cryo} and purged to remove any residual gas; then, the vacuum insulation space between the outer and inner regions was evacuated.

Commercially obtained Kr gas\footnote{Krypton 5.0 from supplier Linde AG with the following maximum components of impurities: \ce{H2O} $\leq$ 2~ppm, \ce{O2} $\leq$ 0.5~ppm, \ce{HC} $\leq$ 0.5~ppm, \ce{N2} $\leq$ 2~ppm, Ar $\leq$ 1~ppm, \ce{CF4} $\leq$ 1~ppm, Xe $\leq$ 1~ppm, CO+\ce{CO2} $\leq$ 1~ppm, \ce{H2} $\leq$ 1~ppm.} 
 was passed once through the filters and then condensed in the LKr vessel.
Measurements of the charge distribution observed at the anode A1 were made as a function of drift time relative to the prompt scintillation light signal in coincidence with the NaI(Tl) signal. 
Only events with single pulses on A1 were used in the off-line analysis to determine the measured energy obtained from the pulse amplitude\footnote{Events with multiple charge deposits within the single pulse width (about 4~\textmu s) were treated as a single pulse. A simulation indicated that these represent only a few percent of detected pulses having a negligible effect on the attenuation measurements.}.
Measurements were made using drift fields $E=0.33$~kV/cm and $E=0.77$~kV/cm.
By measuring the maximum time for ionization electrons to drift between the cathode and grid relative to the MPPC signal times, the drift velocities of $v^{0.33~\mathrm{kV}}_d=1.8 \pm 0.1$~mm/{\textmu}s at 0.33~kV/cm and $v^{0.77~\mathrm{kV}}_d=2.4\pm 0.1$~mm/{\textmu}s at 0.77~kV/cm were obtained; these values are consistent with measurements reported in~\cite{Yoshino:1976zz}.
With no electric field applied to the TPC cathode, the rate of MPPCs hits was found to be approximately 10~kHz\footnote{The detection threshold was estimated to be 100 keV to be compared with the \textsuperscript{85}Kr beta decay endpoint energy of 687 keV.}, due primarily to the presence of \textsuperscript{85}Kr; this rate is consistent with an expected \textsuperscript{85}Kr contamination of $10^{-11}$ to $10^{-10}$~\cite{Kr85}. 

\begin{figure}
\includegraphics[width=\textwidth]{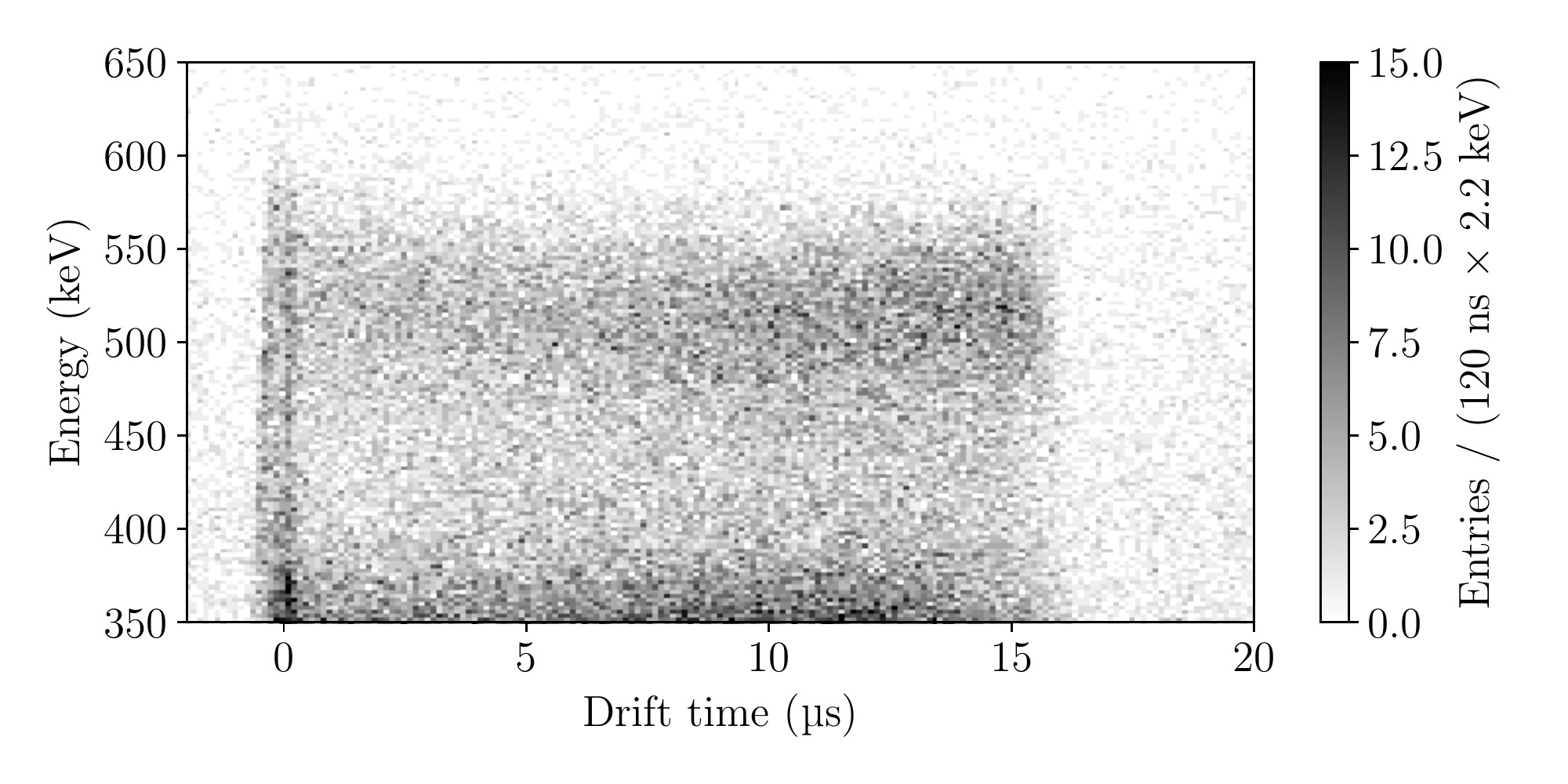}
\caption{Energy distribution from the charge measurement on A1 as a function of the drift time at electric field $E=0.77$~kV/cm. The upper band corresponds to the 511 keV photoelectrons, while the lower energy region is populated by Compton scattering events. The enhancement  near zero drift time corresponds to events with photons converting at the grid. 
Some background caused by accidental \textsuperscript{85}Kr decays in the chamber is evident outside the drift region.
}
\label{EnergyVTime}
\end{figure}

\begin{figure}
\includegraphics[width=\textwidth]{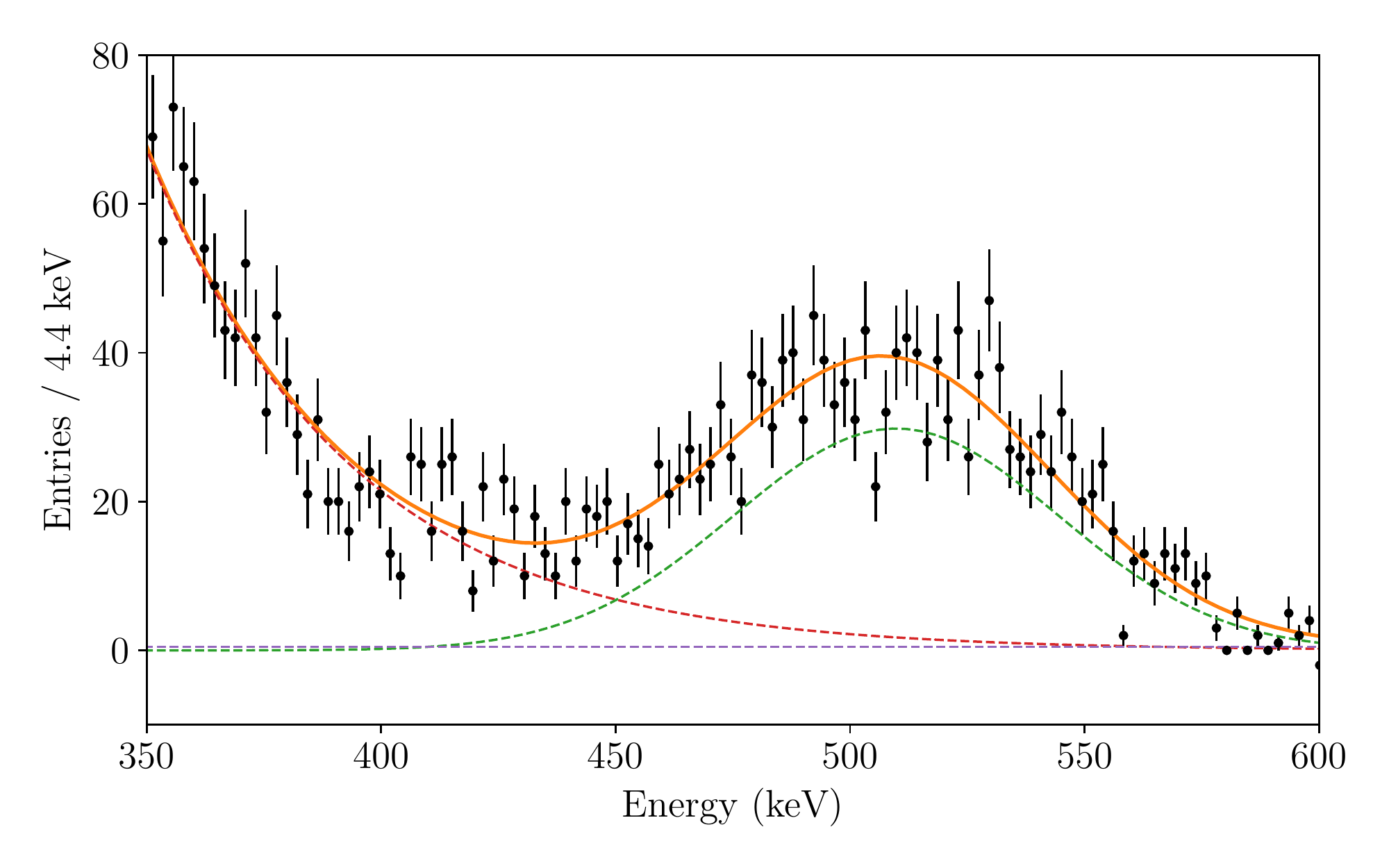}
\caption{Energy distribution obtained by measuring the charge in the central drift time region at an electric field of $E = 0.77$~kV/cm.
The solid orange line represents a fit of the spectrum, which is modeled as a Gaussian 511 keV photoelectron peak (shown as a dashed green line) with exponential (dashed red) and constant (dashed purple) background shapes (more details in the text).}
\label{Charge}
\end{figure}

Figure~\ref{EnergyVTime} shows the distribution of energies based on the charge measurements on A1 vs. the drift time relative to a coincidence of the MPPC and NaI(Tl) at a drift field $E= 0.77$~kV/cm. 
Figure~\ref{Charge} shows the projected energy distribution from Figure~\ref{EnergyVTime} for a central 1~{\textmu}s drift time window.
The fit in the 511 keV  photoelectron peak region was done using a Gaussian function in the presence of  exponential and constant background shape functions.
The resolution for the Gaussian peak was $\sigma_E/E = 6.9 \pm 0.2 $~\%\footnote{The average light signal varies in amplitude due to the solid angle acceptance of the MPPCs which is maximum at the center of the TPC. After selecting the 511 keV photoelectron charge peak, the observed energy resolution based on the light signal alone at the central 1~{\textmu}s drift region was $\sigma_E/E = 10.5\pm 1.1$~$\%$. The charge signal increases and the light signal decreases with increasing electric field\cite{Kubota:1979ugr}. 
Combining the charge and light signals as discussed in \cite{Aprile:2007qd} results in improved resolution $\sigma_E/E = 3.8 \pm 0.2$~$\%$. The correlation angle\cite{Aprile:2007qd} in the charge vs. light plot was approximately 18\degree. In the attenuation analysis described below, only the charge signal is used.}.

To estimate the ionization electron attenuation, the drift region was segmented into 1~{\textmu}s bins and the amplitudes of the 511~keV photoelectron peaks (based on the charge collected by anode A1) were determined in each bin by fitting the  peak regions as described above.
The typical photoelectron peak position uncertainty was approximately $dP=3.6$~keV obtained by adding in quadrature the fit uncertainty of 2~keV and an estimated systematic uncertainty of 3~keV accounting for variations in the fitting region and the background shape. Using simulations of attenuation measurements with peak position uncertainties $dP$ we estimate that lifetimes up to 7~ms could be measured and distinguished from $\infty$ (i.e. no attenuation).

\begin{figure}
\includegraphics[width=\textwidth]{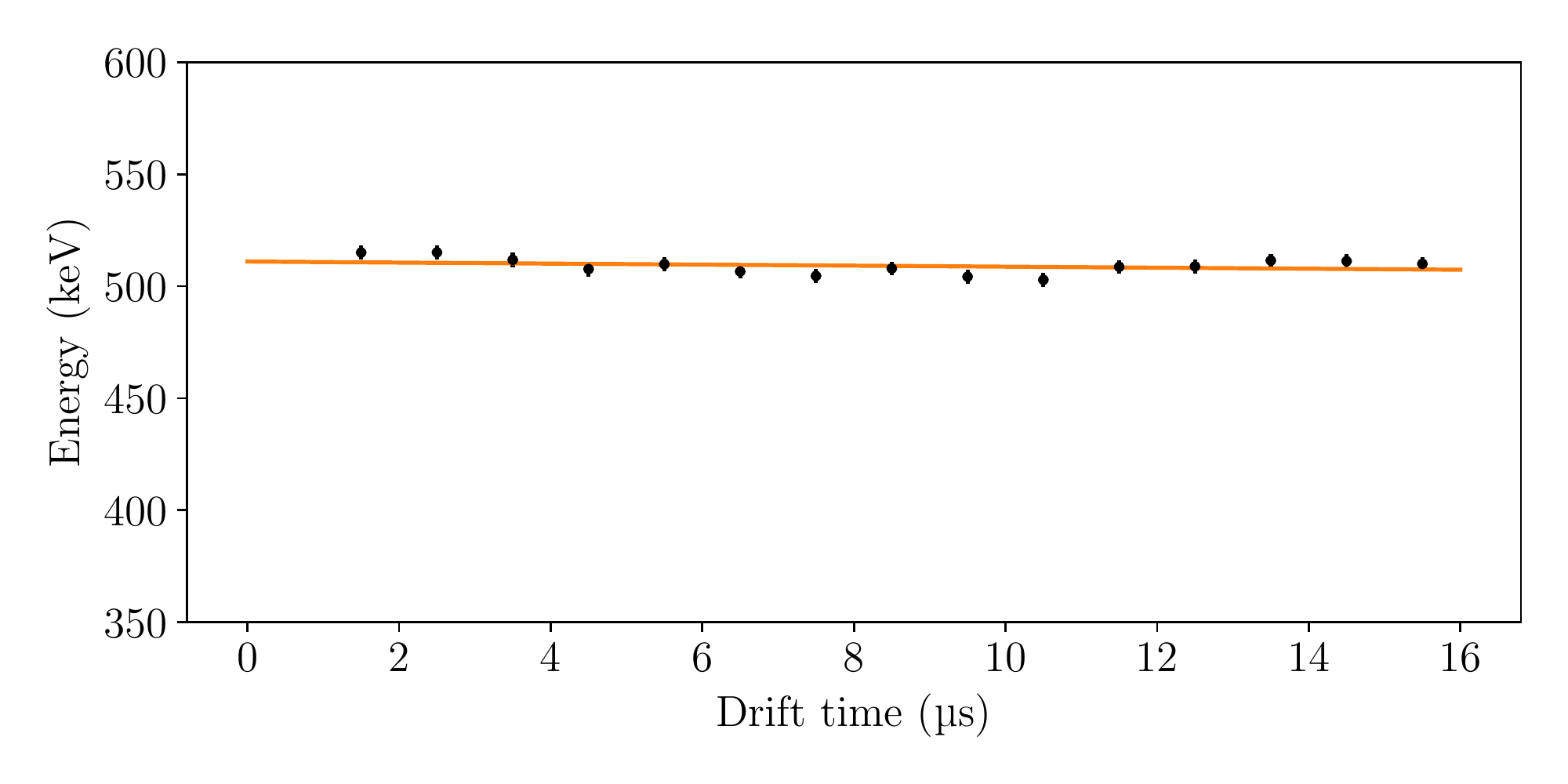}
	\caption{Photoelectron peak energy (keV) from the A1 charge measurement vs. drift time ({\textmu}s) in 1~{\textmu}s slices at electric field $E=0.77$ kV/cm for condensed filtered Kr gas from bottles. The solid orange line shows the fit to an exponential function (see text).}
\label{PeakVTime_bottle}
\end{figure}

Figure~\ref{PeakVTime_bottle} shows the peak positions for the 511~keV photoelectrons as a function of drift time at drift field $E=0.77$~kV/cm based on the A1 charge measurement. 
The photoelectron peak position vs. time distribution was fit to an exponential of the form $A_0 e^{- \lambda t}$ where $A_0$ is normalized to 511 keV at time $t=0 $, defined as the time of drift from the shielding grid to A1, and $\lambda$ is the ionization electron attenuation coefficient; the electron lifetime is $\tau_e=1/\lambda$.
The result of the fit is $\lambda= 449\pm367~ s^{-1}$ with $\chi^2/\mathrm{d.o.f.}= 1.3$.
The 90 \% c.l. upper limit \cite{Feldman:1997qc} is $\lambda<1.04x10^3~ s^{-1}$ which is equivalent to a limit on the ionization electron lifetime $\tau^{min}_e > 0.96 $~ms.
An estimate of the equivalent \ce{O2} (or \ce{H2O}) contamination can be obtained using the average of the atomic electron attachment cross sections available for LAr and LXe compiled by Doke \cite{Doke:1981eac} at this electric field, the measured drift velocity $v^{0.77kV}_d$, and the measured lifetime.
The contamination is roughly estimated to be $<0.2$ ppb. Consistent results were found for drift field $E=0.33$~kV/cm.
Since the attenuation is reduced further at higher fields\cite{Doke:1981eac}, it is clear that this level of impurity contamination has a negligible effect on the operation of the LKRC at its nominal field of $E=3$ kV/cm.

\section{Conclusion}
A system for measuring the attenuation of drifting ionization electrons in the presence of electric fields, and, consequently, the purity of LKr was developed.
The attenuation of drifting ionization was measured in a small-time projection chamber triggered by scintillation light detected by MPPCs in coincidence with signals from a NaI(Tl) crystal observing back-to-back 511~keV annihilation photons from a \textsuperscript{22}Na source. Electron lifetime $\tau_e > 0.96$~ms was observed in gas samples from commercially supplied krypton gas bottles after filtering; similar results were found for gas taken directly from the NA62 LKr calorimeter.
These results indicate that the purity of filtered krypton gas is more than sufficient to satisfy the requirements for operation of the NA62 experiment.

\section*{Acknowledgements}
We thank members of the NA62 collaboration for their assistance. We also thank M. Constable and A. Sorokin for design and implementation of the upgraded TPC mechanical elements. We also thank A. Goncalves Martins de Oliveira for assistance with the setup and Dirk Mergelkuhl for assistance with the alignment. We are also grateful to the CERN Geodetic Metrology Group and the CERN Vacuum Group for their assistance. This work was supported by TRIUMF and NSERC (Canada) grant SAPPJ--2018--0017.
\bibliographystyle{elsarticle-num-names} 
\bibliography{bibliography.bib}

\end{document}